\documentclass[a4paper]{PoS}

\title{ NIKA 2: next-generation continuum/polarized camera at the IRAM 30 m telescope and its prototype}

\ShortTitle{NIKA 2: next-generation continuum/polarized camera at the IRAM 30 m telescope}

\author{\speaker{A.~Ritacco} \thanks{alessia.ritacco@lpsc.in2p3.fr} $^a$, {R.~Adam} $^aj$ , {A.~Adane} $^b$, {P.~Ade} $^c$, {P.~Andr\'e} $^d$, {A.~Beelen} $^e$, {A.~Beno\^it} $^f$, {A.~Bideaud} $^c$, {N.~Billot} $^g$, 
{O.~Bourrion} $^a$, {M.~Calvo} $^f$, {A.~Catalano} $^a$, {G.~Coiffard} $^b$, {B.~Comis} $^a$, {F.-X.~D\'esert} $^h$, {S.~Doyle} $^c$, {J.~Goupy} $^f$, {C.~Kramer} $^g$, 
{S.~Leclercq} $^b$, {J.F.~Mac\'ias-P\'erez} $^a$, {P.~Mauskopf} $^c$, {A. Maury} $^d$, {F.~Mayet} $^a$, {A.~Monfardini} $^f$, {F.~Pajot} $^i$, {E.~Pascale} $^c$, {L.~Perotto} $^a$, {G. Pisano}$^c$, {N.~Ponthieu} $^h$, {M. Rebolo-Iglesias} $^a$ ,
{V.~Rev\'eret} $^d$, {L.~Rodriguez} $^d$, {F.~Ruppin} $^a$, {G.~Savini} $^j$, {K.~Schuster} $^b$, {A.~Sievers} $^g$, {S.~Triqueneaux} $^f$, {C.~Tucker} $^c$, {R.~Zylka} $^b$ \\
\llap {$^a$}Laboratoire de Physique Subatomique et de Cosmologie, Universit\'e Grenoble Alpes, CNRS/IN2P3, 53, avenue des Martyrs, Grenoble, France ; 
 \llap {$^b$}Institut de RadioAstronomie Millim\'etrique (IRAM), Grenoble, France ;
 \llap {$^c$}Astronomy Instrumentation Group, University of Cardiff, UK ;
 \llap {$^d$}Laboratoire AIM, CEA/IRFU, CNRS/INSU, Universit\'e Paris Diderot, CEA-Saclay, 91191 Gif-Sur-Yvette, France ;
 \llap {$^e$}Institut d'Astrophysique Spatiale (IAS), CNRS and Universit\'e Paris Sud, Orsay, France ;
 \llap {$^f$}Institut N\'eel, CNRS and Universit\'e de Grenoble, France ;
 \llap {$^g$}Institut de RadioAstronomie Millimetrique (IRAM), Granada, Spain ;
 \llap {$^h$}Institut de Plan\'etologie et d'Astrophysique de Grenoble (IPAG), CNRS and Universit\'e Grenoble Alpes, France ;
 \llap {$^i$}University College London, Department of Physics and Astronomy, Gower Street, London WC1E 6BT, UK;
 \llap {$^j$} Laboratoire Lagrange, Université Côte d'Azur, Observatoire de la Côte d'Azur, CNRS, Blvd de l'Observatoire, CS 34229, 06304 Nice cedex 4, France
}

\FullConference{EXTRA-RADSUR2015 (*)\\
		20--23 October 2015\\
		Bologna, Italy

                \bigskip
                \hrule
                \bigskip

                \textnormal{(*) This conference has been organized
                  with the support of the Ministry of Foreign Affairs
                  and International Cooperation, Directorate General
                  for the Country Promotion (Bilateral Grant Agreement
                  ZA14GR02 - Mapping the Universe on the Pathway to
                  SKA)}
}

\abstract{{\it NIKA 2} (New Instrument of Kids Array) is a next generation continuum and polarized instrument successfully installed in October 2015 at the IRAM 30 m telescope on Pico-Veleta (Granada, Spain).\
 {\it NIKA 2} is a high resolution dual-band camera, operating with frequency multiplexed LEKIDs (Lumped Element Kinetic Inductance Detectors) cooled at 100 mK. 
 Dual color images are obtained thanks to the simultaneous readout of a 1020 pixels array at 2 mm and 1140 x 2 pixels arrays at 1.15 mm with a final resolution of 18 and 12 arcsec respectively, and 6.5 arcmin of Field of View (FoV). The two arrays at 1.15 mm allow us to measure the linear polarization of the incoming light. This will place {\it NIKA 2} as an instrument of choice to study the role of magnetic fields in the star formation process. The {\it NIKA} experiment, a prototype for {\it NIKA 2} with a reduced number of detectors ($\sim$ 400 LEKIDs) and FoV (1.8 arcmin), has been successfully operated at the IRAM 30 telescope in several open observational campaigns. The performance of the {\it NIKA 2} polarization setup has been successfully validated with the {\it NIKA} prototype.
}

\begin{document}

\section{Introduction}
Magnetic fields play a crucial role in the dynamics of many astrophysical processes including star formation, mediating shocks, influencing heat and mass transport and cosmic rays \cite{crutcher}. 
Mapping observations of linearly-polarized continuum emission resulting from magnetically-aligned dust grains at mm and submm wavelengths is a powerful tool to measure the morphology and structure of magnetic field lines in star-forming clouds and dense cores \cite{Matthews2002}. High angular resolution observations of polarization provided by {\it NIKA 2} instrument will improve our understanding of the star formation physics on galactic scales, still poorly constrained observationally.  

{\it NIKA 2} is a dual band camera \cite{calvo2016} installed in October 2015 at the IRAM 30 m telescope on Pico Veleta (Spain). {\it NIKA 2} consists of three arrays of Kinetic Inductance Detectors observing the sky at two frequency bands: 260 (1140 x 2 KIDs) and 150 (1020 KIDs) GHz.
{\it NIKA 2} has a resolution of 12 and 18 arcsec at 260 GHz and 150 GHz respectively and a FoV of 6.5 arcmin.

Its prototype, the {\it NIKA} camera, with a reduced number of pixels ($\sim$ 400) and FoV ($\sim$ 1.8 arcmin) has demonstrated its scientific competitiveness via intensity observations of various astrophysical sources \cite{catalano2014} and in particular of the Sunyaev-Zel'dovich effect on clusters of galaxies \cite{adam2014, adam2015}.
In the following we discuss the performance and calibration of the polarization setup used on {\it NIKA} and chosen as a solution for the polarimeter of the {\it NIKA 2} instrument.

\section{NIKA and NIKA 2 polarization system}
The state-of-the-art dual band camera {\it NIKA}, operating at the IRAM 30 m telescope until February 2015, consisted of two KIDs arrays cooled down by a dilution cryostat allowing to reach the detector optimal working temperature of 100 mK. It observed the sky in intensity and polarization in two frequency bands with central frequency of 150 and 260 GHz. 

\begin{figure}
\begin{center}
\includegraphics[width=.8\textwidth]{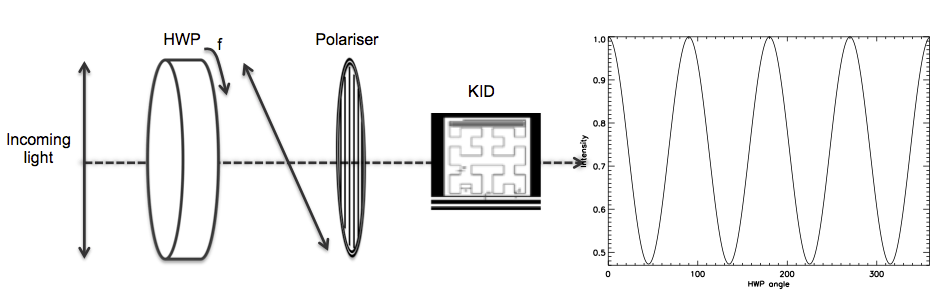}
\caption{Schematic view of the observational polarization setup. From left to right we observe the rotating multi-layer HWP, the polarizer and a  Hilbert geometry KID element of the matrices mounted in the {\it NIKA} cryostat. }\label{fig1}
\end{center}
\end{figure}

The small time constant of {\it NIKA} detectors LEKIDs \cite{monfardini2010,doyle2008,roesch} permits the use of fast modulation techniques to shift the polarized signal to high frequency far away from the dominant atmospheric contamination at low frequencies. {\it NIKA} polarization system consists of a warm rotating multi-layer Half Wave Plate (HWP hereafter) plus a polarizer (which is necessary because the {\it NIKA} detectors are sensitive to both polarizations \cite{roesch}), and the {\it NIKA} instrument (cryostat and KIDs matrices). For more details on the {\it NIKA} instrument see \cite{monfardini2011, catalano2014}.
{\it NIKA} polarization setup is shown in Fig. \ref{fig1}.
A similar polarization system has been integrated in the {\it NIKA 2} instrument, which has the rotating HWP mounted on the pupil of the cryostat and a polarizer mounted at the 100 mK stage to split the two components of the linear polarization on two matrices of 1140 pixels operating at 260 GHz.

The {\it NIKA} system polarization efficiency characterization discussed in \cite{2015JLTP..tmp...76R} shows a polarized light quasi totally transmitted. In particular, considering an ideal polarizer, the estimation of the HWP transmission coefficients gives an estimation of the polarization efficiency $\rho_{pol}$ of the system. We find $\rho_{pol}$ $\simeq$ 0.994 and 0.986 at 1 and 2 mm respectively, with uncertainties below 1$\%$. 

\newpage
\section{Observation strategy and results}
The polarized input signal is modulated in time to four times the frequency of the HWP ($\sim$ 3 Hz). 
Thus, the polarized astrophysical signal is located far from the atmospherical emission dominating at lower frequencies with a $1/f$ like spectrum. Imperfections of the HWP modulate the background and lead to additional signal on the timelines, peaking at harmonics of the HWP rotation frequency. This parasitic signal is removed in the data analysis pipeline.
The data output timeline for a KID $k$ is described by:

\begin{eqnarray}
 m_{k} &=& \frac{1}{2}\{I + {\rho}_{\rm pol}[Q\cos(4{\omega}t + 2{\alpha}_{\rm Sky}(p_{\rm t})) + Usin({4{\omega}t} \nonumber 
 	  + 2{\alpha}_{Sky}(p_{\rm t}))]+ S_{\rm parasitic} ({\omega}t, 2{\omega}t, 3{\omega}t, 4{\omega}t ...)\} \nonumber \\
	  &+& {\rm noise}_{\rm atmospheric} + {\rm noise}_{\rm detector}
 \label{pol_eq}
 \end{eqnarray}
 where $p_{\rm t}$ represents the pixel observed at time $t$ and ${\alpha}_{\rm Sky}$ the angle between the telescope reference frame and the local meridian on the sky. 
The angle ${\alpha}_{\rm Sky}$ is measured from north to east in the equatorial system as  ${\alpha}_{\rm Sky} =- \{{\epsilon} + {\eta}$\},
where ${\epsilon}$ represents the elevation, ${\eta}$ the parallactic angle.

A specific polarization data analysis pipeline has been developed. Briefly, $I, Q, U$ Stokes parameters maps in sky coordinates are constructed by applying the following procedures: i) construction of a template of the HWP modulation and removal of parasitic signal; ii) construction of $Q$ and $U$ TOIs (time ordered informations) from the $4\omega t$ component; iii) removal of atmospheric noise both in intensity and polarization; iv) and, projection of the $I, Q, U$ TOIs into $I, Q, U$ maps. 

Observations on Uranus, considered to be unpolarized, showed an intensity to polarization leakage effect \cite{2015JLTP..tmp...76R}. For point sources we can correct for this effect using the observed Uranus intensity to polarisation leakage as a template. However, for extended sources a more complex algorithm has been developed. 
The residual instrumental polarization after correction drops from 3\% of the total intensity to below 0.1\%.
We observed several known quasars with low variability in time and known extended sources to calibrate the polarization orientation. 
Measured polarization angle for the quasar 3C 286 \cite{perley&butler}, {$\alpha_{Sky}$ = [29.0 $\pm$ 3.9]$^\circ$ and $\alpha_{Sky}$ = [27.7 $\pm$ 3.9]$^\circ$ at 260 GHz and 150 GHz, respectively, agrees with the literature \cite{xpol}.

We show here the observation of the Crab nebula (Tau A, M1 or NGC 1952) which is a well known polarization calibrator at mm wavelengths \cite{macias} . The Crab nebula is a supernova remnant that emits a highly polarized signal due both to the synchrotron emission of the central pulsar and its interaction with the surrounding gas. 
In Fig. \ref{fig2} we show the $I, Q, U$ maps of the Crab nebula observed at 150 GHz \cite{2015JLTP..tmp...76R}, the results are in a good agreement with previous observations using the XPOL/EMIR at the IRAM 30 m telescope \cite{aumont}.

\begin{figure}
\begin{center}
\includegraphics[width=.3\textwidth]{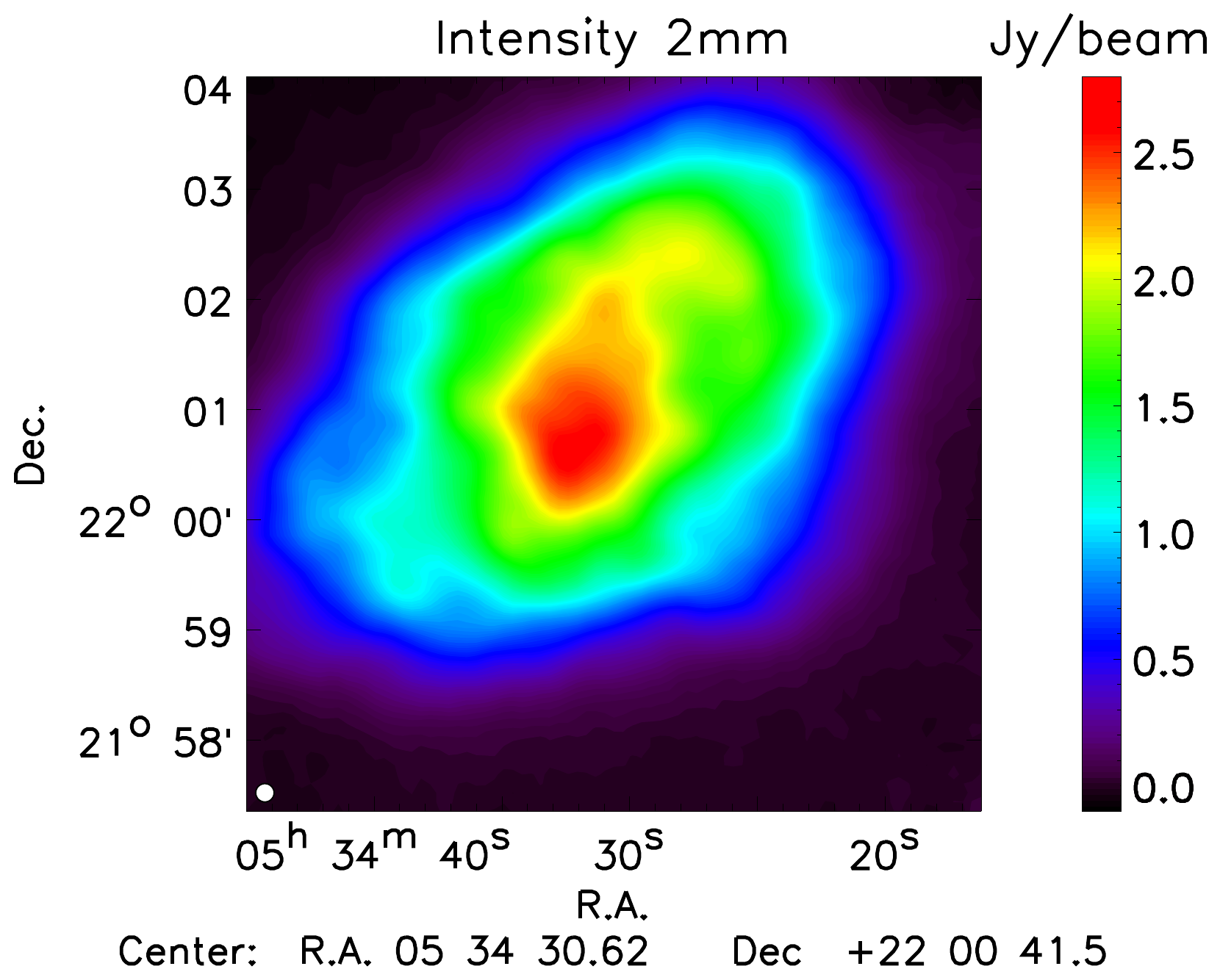}
\includegraphics[width=.3\textwidth]{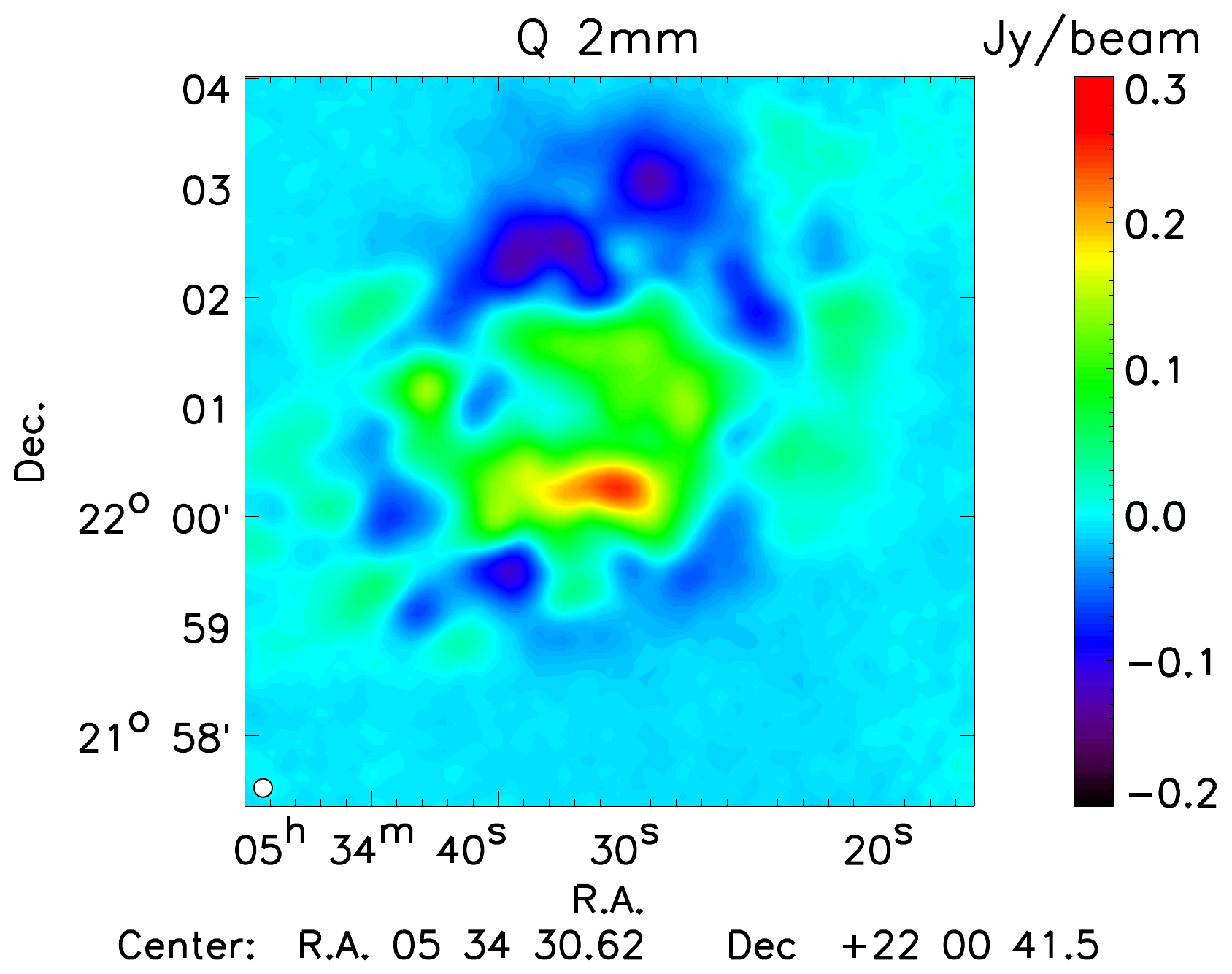}
\includegraphics[width=.3\textwidth]{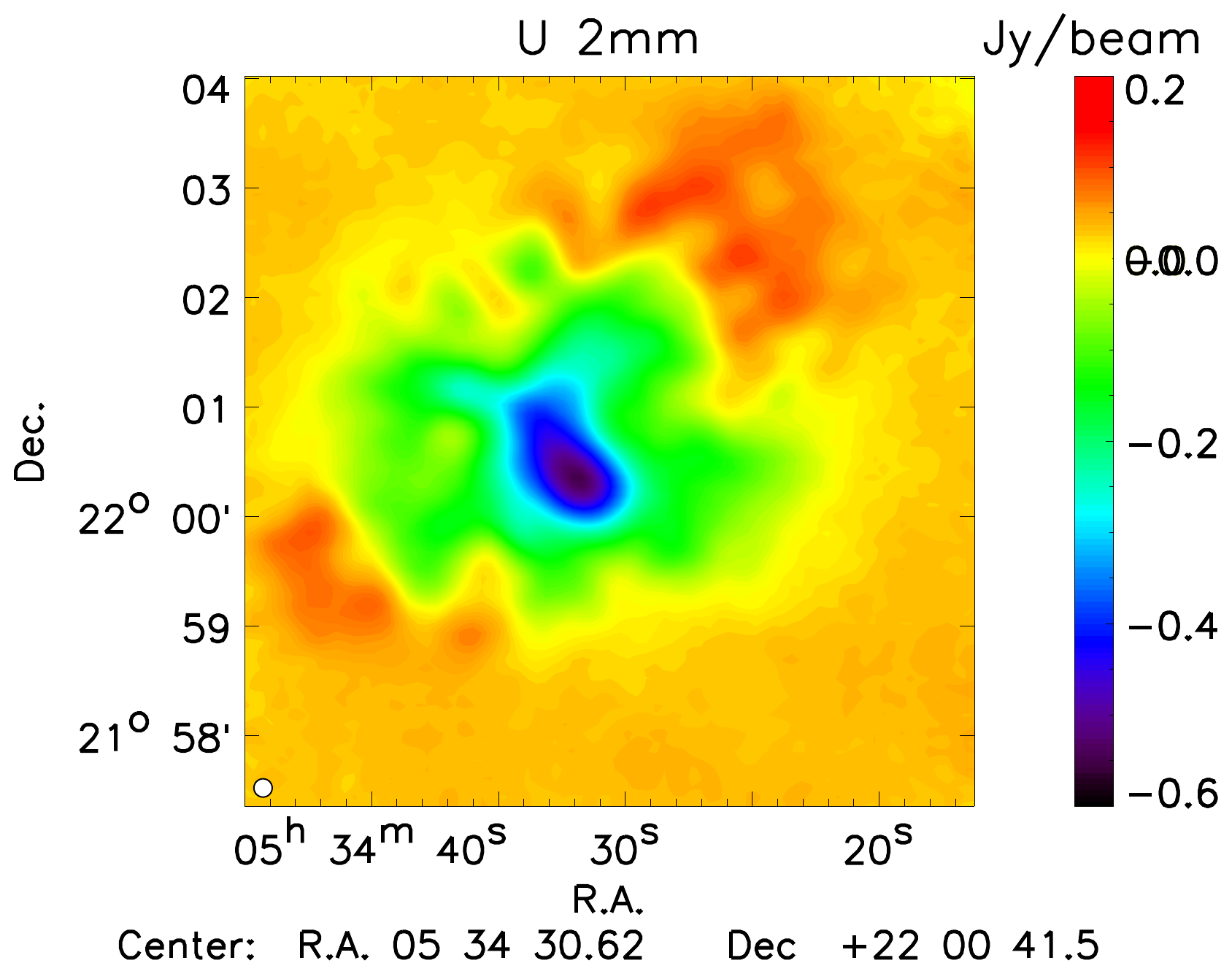}
\caption{Stokes parameters $I, Q, U$ maps at 150 GHz of the Crab nebula.}
\label{fig2}
\end{center}
\end{figure}

\section{Conclusion}
The {\it NIKA} polarization module consists of a rotating HWP plus a polarizer placed on the pupil of the cryostat at the IRAM 30 m telescope. It has been proven to be efficient to measure the polarization of celestial sources, i.e. quasars and the Crab nebula. A similar polarization setup is mounted on {\it NIKA 2}. {\it NIKA 2} has been mounted at the IRAM 30 m telescope in October 2015, the three arrays with $\sim$ 3300 KIDs have been successfully installed, more than 80$\%$ are valids. In winter  2015/2016 and in summer 2016 the camera will be in commissioning phase, opening to all IRAM users is expected to take place during the winter semester 2016/17. Since the scientific results obtained with the {\it NIKA} camera are very promising, we expect that {\it NIKA 2} becomes an avant-garde imaging instrument at millimeter wavelengths.

\acknowledgments{\footnotesize We would like to thank the IRAM staff for their support during the campaign. 
This work has been partially funded by the Foundation Nanoscience Grenoble, the ANR under the contracts "MKIDS" and "NIKA". 
This work has been partially supported by the LabEx FOCUS ANR-11-LABX-0013. 
This work has benefited from the support of the European Research Council Advanced Grant ORISTARS under the European Union's Seventh Framework Programme (Grant Agreement no. 291294).
The NIKA dilution cryostat has been designed and built at the Institut N\'eel. In particular, we acknowledge the crucial contribution of the Cryogenics Group, and in particular Gregory Garde, Henri Rodenas, Jean Paul Leggeri, Philippe Camus. 
R. A. would like to thank the ENIGMASS French LabEx for funding this work. R.A. acknowledges support from the CNES post-doctoral fellowship program.
A. R. would like to thank the FOCUS French LabEx doctoral fellowship program. 
A. R. acknowledges support from the CNES doctoral fellowship program.
}

\bibliography{bibliography}

\end{document}